\documentclass[a4paper, notitlepage, 12pt]{article}
\usepackage{geometry}
\fontfamily{times}
\usepackage{setspace,relsize}               
\usepackage{moreverb}                        
\usepackage{xurl}
\usepackage{hyperref}
\hypersetup{colorlinks=true,citecolor=magenta}
\usepackage{amsmath}
\usepackage{mathtools} 
\usepackage{amsthm}
\usepackage{amssymb}
\usepackage{indentfirst}
\usepackage{todonotes}
\usepackage[authoryear,round]{natbib}
\usepackage[pdftex]{lscape}
\usepackage[utf8]{inputenc}
\usepackage{algpseudocode}
\usepackage{booktabs}
\usepackage{multirow}
\usepackage[T1]{fontenc}
\usepackage{algorithm}
\usepackage{algpseudocode}

\usepackage{libertine}
\usepackage[libertine]{newtxmath}

\usepackage{xcolor}
\usepackage{pagecolor}
\usepackage{lipsum}

\usepackage{array}
\usepackage{longtable}
\usepackage{epsfig}
\usepackage{graphicx}
\usepackage{graphics}
\usepackage{float}
\usepackage{subfigure}
\usepackage{multirow}
\usepackage{color}
\usepackage{fullpage}
\usepackage[normalem]{ulem} 
\usepackage{makeidx}
\usepackage{xspace}
\usepackage{wrapfig}
\usepackage{booktabs}
\usepackage{url}
\usepackage{subcaption}

\usepackage [autostyle, english = british]{csquotes}
\MakeOuterQuote{"}

\title{\vspace{-9ex}\centering \bf Large-scale Epidemiological modeling: Scanning for Mosquito-Borne Diseases Spatio-temporal Patterns in Brazil}
\author{
Eduardo C. Araujo$^{1}$, Claudia T. Codeço$^{2}$, Sandro Loch$^{1}$, \\
Luã B. Vacaro$^{1}$, Laís P. Freitas$^{3, 4}$, Raquel M. Lana$^{5}$,\\
Leonardo S. Bastos$^{2}$, Iasmim F. de Almeida$^{6}$, Fernanda Valente$^{7}$,\\
Luiz M. Carvalho$^{1}$ and Flávio C. Coelho$^{1}$\\
$^{1}$\footnotesize{EMAp - FGV, Brazil},
$^{2}$\footnotesize{PROCC-FIOCRUZ,
Brazil},
$^{3}$\footnotesize{ESPUM, Canada},\\
$^{4}$\footnotesize{CReSP, Canada},
$^{5}$\footnotesize{BSC, Spain}, 
$^{6}$\footnotesize{ENSP-FIOCRUZ, Brazil},
$^{7}$\footnotesize{Observatório de Bioeconomia, FGV, Brazil}
}

\begin{document}
\maketitle

\begin{abstract}
The influence of climate on mosquito-borne diseases like dengue and chikungunya is well-established, but comprehensively tracking long-term spatial and temporal trends across large areas has been hindered by fragmented data and limited analysis tools.
This study presents an unprecedented analysis, in terms of breadth, estimating the SIR transmission parameters from incidence data in all 5,570 municipalities in Brazil over 14 years (2010-2023) for both dengue and chikungunya.
We describe the Episcanner computational pipeline, developed to estimate these parameters, producing a reusable dataset describing all dengue and chikungunya epidemics that have taken place in this period, in Brazil.
The analysis reveals new insights into the climate-epidemic nexus: We identify distinct geographical and temporal patterns of arbovirus disease incidence across Brazil, highlighting how climatic factors like temperature and precipitation influence the timing and intensity of dengue and chikungunya epidemics.
The innovative Episcanner tool empowers researchers and public health officials to explore these patterns in detail, facilitating targeted interventions and risk assessments.
This research offers a new perspective on the long-term dynamics of climate-driven mosquito-borne diseases and their geographical specificities linked to the effects of global temperature fluctuations such as those captured by the ENSO index. 

\end{abstract}

\section{Introduction}

Dengue and chikungunya are among the most significant mosquito-borne diseases in terms of disease burden and potential for global expansion. Worldwide, the estimated number of dengue cases has increased from 30 million, in 1990, to 56 million in 2019 \citep{Yang21}, reaching more than 100 countries. Chikungunya, primarily transmitted by the same mosquitoes \textit{Aedes aegypti} and \textit{Aedes albopictus}, has been spreading worldwide since the early 2000s and is now present in all continents with tropical zones \citep{Manzoor22}.   

The temporal dynamics of these vector-borne diseases is characterized by seasonal and multi-year cycles that can vary from place to place. Within seasons, they can further differ in their reproductive number, peak size, timing and total attack rate \citep{almeida_2022}.
This spatial-temporal heterogeneity is strongly associated with the climate and meteorological conditions that affect the mosquito vectorial capacity and viral transmission, in combination with the history of previous population exposure that affects the level of collective immunity. The urban environment, characterized by crowding, inadequate garbage and water services, can further increase the population vulnerability and exposure, while effective vector control measures can potentially flatten the curve. As a result, all of these local factors contribute to generate very diverse spatially-temporal epidemic dynamics locally.    

Dengue has maintained a continuous presence in the tropical regions of Brazil since 1986~\citep{rodriguez2011re}, but with patterns varying from transient transmission to epidemic and endemic persistence \citep{almeida_2022}.
Chikungunya fever was introduced in Brazil in 2014 and persisted, with outbreaks increasing in frequency \citep{de2023spatiotemporal}.  In the last years, probably due to the gradual warming of the southern states, outbreaks of both diseases have begun to occur in areas with subtropical and temperate climates \citep{Codeco_Oliveira_Ferreira_Riback_Bastos_Lana_Almeida_Godinho_Cruz_Coelho_2022, de_Almeida_2023}. Understanding how climate can affect the reproductive number and timing of dengue and chikungunya epidemics locally is useful for guiding decisions at the national level.  

This study is centred on estimating and examining the characteristics of all the dengue and chikungunya epidemics that occurred in Brazilian municipalities from 2010 to 2023. The key descriptors are the reproductive number, size and timing, estimated by the Episcanner pipeline, a computational pipeline developed for this end. 

\section{Methods}

\subsection{Data Sources}
\label{sec:data}
Brazil comprises 5570 municipalities, divided into 26 states, plus the Federal District, the country's capital. The disease data for all the municipalities was obtained from the Infodengue project~\citep{codeco_infodengue_2018}, available at \url{info.dengue.mat.br}, for the period between 2010 and 2023. 
The Infodengue project gathers notification data for dengue, chikungunya, and Zika from the Brazilian Ministry of Health, sanitizes and re-publishes it, along with epidemiological analyses, such as estimates for the effective reproduction number, ${\cal R}_t$, by week, for all municipalities.
The code for obtaining the Infodengue data is available on the GitHub repository associated with this paper.
The climate data were obtained from the Copernicus ERA5 reanalysis dataset~\citep{Copernicus_Muñoz_Sabater_2019} and daily averaged by the municipality. The climate variables included land-surface temperature, relative humidity, precipitation and atmospheric pressure.

\subsection{The Episcanner Pipeline}

If all municipalities experienced a dengue epidemic per year, in 10 years, there would be 55,700 epidemics, varying in velocity, timing, and magnitude. This is a large number of models to be fitted individually. Episcanner is a computational pipeline that efficiently scans all these potential epidemics and estimates the epidemiological descriptors whenever an epidemic is detected. The epidemic descriptors are the basic reproduction number (${\cal R}_0$), peak week, the total outbreak size, as well as timing and duration of the epidemic.

The core of the Episcanner pipeline is the Richards logistic growth function, a sigmoid curve that has a one-to-one relationship with the SIR epidemiological model \citep{wang2012richards}. Based on this relationship, the epidemiologically meaningful parameters of the SIR (Susceptible-Infectious-Recovered) transmission model can be efficiently derived from the fitted Richard model, which serves as a curve template for identifying an epidemic. This simplification is pivotal for curtailing the computational cost associated with estimating the epidemic descriptors of interest for all municipalities and epidemic years considered.

\begin{figure}[!h]
\centering\includegraphics[width=3in]{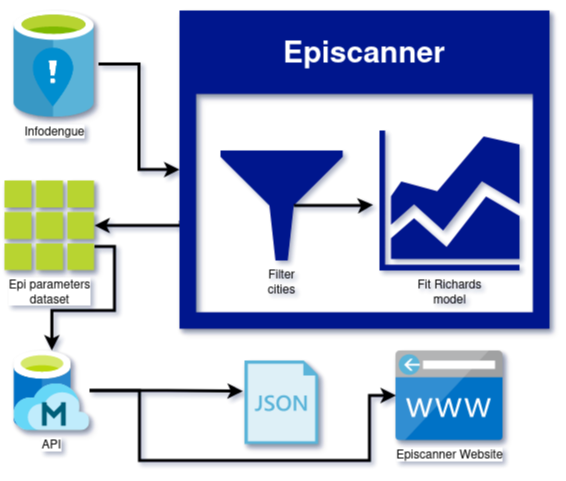}
\caption{The computational pipeline of  Episcanner inferential engine is the dark blue block. Data from Infodengue, after being filtered to identify candidates for epidemic years, is then fitted to the Richards model. The resulting estimated parameter dataset is made available through an API as a JSON object (\url{https://api.mosqlimate.org/datastore/}). This data can be visualized in the Episcanner dashboard, which is part of the Infodengue website (\url{https://info.dengue.mat.br/epi-scanner}).}
\label{fig:workflow}
\end{figure}

The Episcanner pipeline is depicted in Figure \ref{fig:workflow}, and can be summarized in the following steps:
\begin{enumerate}
\item Infodengue's incidence time series are obtained from their API.
\item Filter years with a potential \textit{epidemic}, defined as an annual time series having a minimum of 3 weeks with at least 0,9 probability of ${\cal R}_t>1$ and more than fifty cumulative cases. Any year with less than that is discarded as a short outbreak. This is done for all municipalities. 
\item Split all-time series at week 45 to obtain a collection of 52 week-long time series, one for each year and each municipality. The cut point defines the typical beginning of the dengue season in Brazil.
\item Richards model is fitted to the observed curves, using an optimization algorithm, generating the SIR parameter estimates (see below for details).
\item Epidemiological parameters dataset is created and made available through the Mosqlimate API.
\end{enumerate}

The complete set of parameters produced from all Brazilian municipalities can be browsed through the online dashboard~\citep{Realtime_Epi_Report} called Episcanner, within the Infodengue website.
This dashboard is updated weekly and allows anyone to browse the estimated epidemic characteristics of dengue and chikungunya, as described in this paper.

\subsection{Estimation of the Epidemiological Parameters}
\label{sec:epi_estimation}

The Richards model is defined  by the following equation:
\begin{align}\label{eq:richards}
\begin{split}
 J(t) =  L - L[1 + \alpha e^{b(t-t_j)}]^{-1/ \alpha},   
\end{split}
\end{align}
where $J(t)$ is interpreted as the accumulated number of cases at week $t$, $L > 0$ is the estimated total number of cases at the end of the epidemic, $t_j \in [1,52]$ is the week of the inflexion point of the sigmoid curve, which corresponds to the peak of the epidemic. The remaining parameters $\alpha$ and $b$ are additional parameters to be estimated.

When mapping to the SIR model, the function $J(t)$ corresponds to the number of recovered individuals at time $t$, which in the SIR model is denoted $R(t)$. 
 The equivalent SIR model's parameters can be derived from the Richards equation's parameters (\ref{eq:richards}) through the equations given below (refer to \cite{wang2012richards} for derivation):

\begin{align}\label{beta}
\begin{split}
    \beta = \cfrac{b}{ \alpha },
\end{split}
\end{align}

\begin{align}\label{gamma}
\begin{split}
    \gamma = b \left(\cfrac{1}{\alpha} - 1 \right),  
\end{split}
\end{align}

\begin{align}\label{R0}
\begin{split}
     R_0 = \cfrac{\beta}{\gamma} = \cfrac{1}{1-\alpha}.
\end{split}
\end{align}

\begin{figure}[!h]
\centering\includegraphics[width=5in]{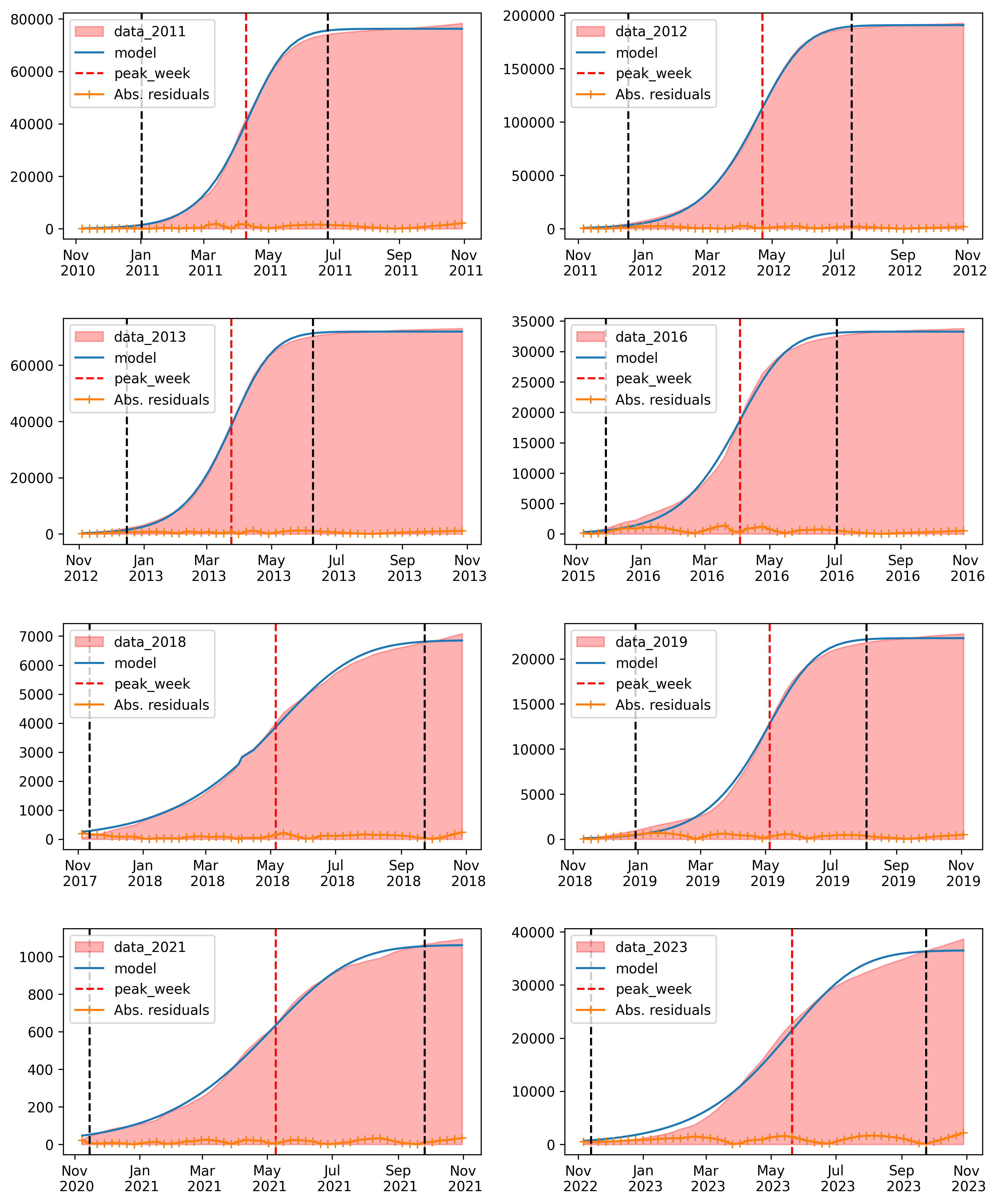}
\caption{Example of the outcome of estimating the start and end weeks of an epidemic for Rio de Janeiro municipality in all epidemic years between 2010 and 2023. The red area represents the cumulative number of cases. Vertical dashed lines indicate the start and end of the epidemic. The blue curve is the fitted Richards curve, and the orange one at the bottom is the absolute error of the fit every week. }
\label{fig:start_end}
\end{figure}

Besides the reproductive number ($R_0$), epidemic size ($L$) and peak week ($t_j$), the Episcanner dataset also includes the onset of the epidemic ($w_s$) that is defined as the week in which the new cases crossed the 5\% percentile, and the final week of the epidemic ($w_e$) defined as the week when the cases dropped below this threshold (See Figure \ref{fig:start_end}). From these points, the epidemic duration ($w_e - w_s$) in weeks is computed and also added to the dataset.

From now on, we will refer to these derived parameters, collectively, as \textit{epidemic parameters} or \textit{epidemic descriptors}.

\subsection{Fitting the model to data}
\label{sec:opt_fit}

For every selected city and epidemic year, as defined in section \ref{sec:data}, we fitted the Richards model (eq. \ref{eq:richards}) to the data by solving the following optimization problem: determining the values of the parameters set $\xi := \{L, \alpha,b, t_j \mid L, \alpha,b, t_j\} \in \mathbb{R}^+$ that minimizes the error function below:
\begin{align}\label{eq:erro}
\begin{split}
	\operatorname{argmin}_{\xi} \sum_{t=1}^{52} \frac{(C(t)-J(t,\xi))^2}{52},
 \end{split}
\end{align}
where $C(t)$ is the observed cumulative number of cases at week $t$.
The following ranges restricted the possible values of the parameters of $\xi$: $\gamma \in [0.95, 1.05]$ (1-week infectious period), $\alpha \in [0.001, 1]$, $b \in [10^{-6}, 1]$, $t_j \in [5, 35]$ (based on the overall distribution of peak weeks).
Additionally, we restricted the possible values of $\alpha$ to always satisfy $\alpha=\frac{b}{\gamma+b}$~\citep{wang2012richards}.

This fitting process is computationally more efficient than to fit the SIR ordinary differential equations, since it avoids numerical integration steps.
The optimization was implemented with the \emph{lmfit} Python package\footnote{\url{https://lmfit.github.io/lmfit-py/intro.html}}, using a global optimization approach based on the differential evolution algorithm~\citep{storn1997differential}. Using the estimated epidemiological parameters, we explored their association with climate variables across time and space.

\subsection{Predicting the week of epidemic peak}
\label{sec:climate_impact}
One question of interest is how early the epidemic peak will occur in a given year, in a given place. 
To test if the dengue epidemic descriptors, estimated by Episcanner, show a correlation to epidemiological, demographic and climate variables, we built a Histogram Gradient Boosting Regression (HGBR) model. The outcome of the model is the peak week of epidemics, $W^{peak}_{i,y}$for every epidemic year in every location. 
Predictors included lagged climate variables, case counts and population sizes, as well as lagged epidemic descriptors. 
The regression model proposed can be written as below
\begin{align}
\begin{split}
    W^{peak}_{i,y} \sim C_{i,y-1}+D_{i,y-1}+E_{i,y-1}+\varepsilon, 
    \label{eq:HGB}
\end{split}
\end{align}
where $W^{peak}_{i,y}$ is the peak week of the epidemic at municipality $i$ on year $y$. $C_{i,y-1}$, $D_{i,y-1}$ and $E_{i,y-1}$ are respective sets of climate, demographic, and epidemiological predictors. A comprehensive description of all model features is provided in Table S1 of the supplementary material.
We used the Scikit-learn Python package to fit the model (\url{https://scikit-learn.org}).

The model was fitted using data encompassing all Brazilian cities for the entire period, filtered as described above for epidemic parameter estimation. A single model was fitted to  each geographical region of Brazil. Brazil has 5 geographical regions, namely: north (N), northeast (NE), midwest (MW), southeast (SE) and south (S). This model was fitted only for dengue due to its broader geographic coverage.
We used Shapley Additive Explanations (SHAP values) framework~\citep{lundberg2017unified} to analyze the importance of each feature (predictor) of the model.

\section{Results}
\label{sec:results}

\subsection{Spatio-temporal patterns of dengue and chikungunya epidemics}

Out of the 5570 Brazilian cities, 4096 (73\%) had at least one epidemic year for dengue and 1263 cities (22\%) for chikungunya between 2010 and 2023.
In total, there were 18567 dengue epidemic events and 2199 chikungunya epidemic events. Table \ref{tab:stats} contains summary statistics for some of the parameters estimated. The full set of parameters also includes the SIR model's $\beta$ and $\gamma$ parameters, the $\alpha$ parameter from the Richards equation, and the starting and ending weeks of the epidemic.

\begin{table}[!h]
\caption{Summary statistics for selected parameters estimated for dengue and chikungunya across all epidemics.}
\label{tab:stats}
\begin{tabular}{lrrrrrr}
\hline
 \textbf{dengue} & ${\cal R}_0$ & duration & peak week & total cases &  incidence (per 100k) \\
\hline
mean & 1.97 & 26.62 & 22.04 & 992.09 &  2146.22 \\
std & 0.52 & 10.38 & 5.80 & 4262.09 &  2841.39 \\
25\% & 1.60 & 19.00 & 18.40 & 113.75 &  523.45 \\
50\% & 1.86 & 25.00 & 22.11 & 240.40 &  1203.41 \\
75\% & 2.20 & 33.00 & 25.24 & 624.70 &  2626.13 \\
\hline
\textbf{chikungunya} &&&&&\\
\hline
mean & 2.12 & 25.47 & 24.69 & 624.43  & 1156.76 \\
std & 0.70 & 12.17 & 7.29 & 2341.62  & 1620.65 \\
25\% & 1.60 & 16.0 & 19.78 & 98.65  & 213.80 \\
50\% & 1.94 & 24.0 & 24.68 & 195.33  & 594.58 \\
75\% & 2.45 & 34.0 & 29.51 & 461.99  & 1436.29 \\
\hline
\end{tabular}
\end{table}

\begin{figure}[!h]
\centering\includegraphics[width=5in]{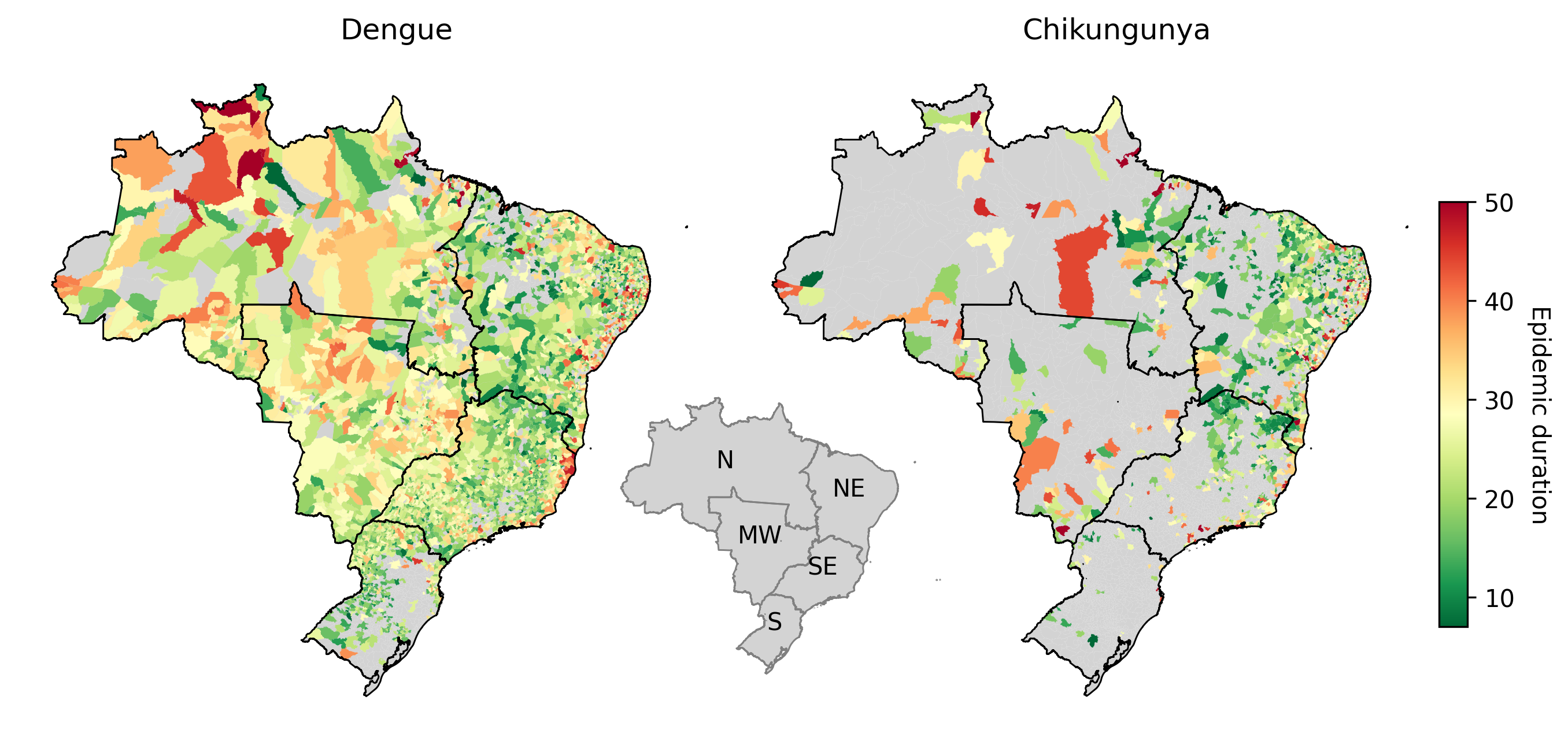}
\caption{Median Epidemic duration (in weeks) estimated for Dengue and Chikungunya cases. Note the increase in the duration of dengue epidemics as one moves from the southern to the northern states.
 This trend is less marked but also visible for chikungunya epidemics. In the smaller map in the middle, we have the geographic regions of Brazil: North(N), Northeast (NE), Midwest (MW), Southeast (SE), and South (S).}
\label{fig:map_duration}
\end{figure}

The spatial distribution of epidemic parameters (${\cal R}_0$, peak week, and epidemic duration) exhibit discernible patterns across different regions of Brazil. 
Notably, the average duration of dengue epidemics exhibits an ascending gradient from the southeast (SE) to the northwest (NW) regions, see Figure \ref{fig:map_duration}.
The reproductive numbers of both dengue and chikungunya exhibited spatial and temporal variability across the country (see Figure S1 of the supplementary material). 
This variability is evident not only between different geographic regions but also over time.

\begin{figure}[!h]
\centering\includegraphics[width=5in]{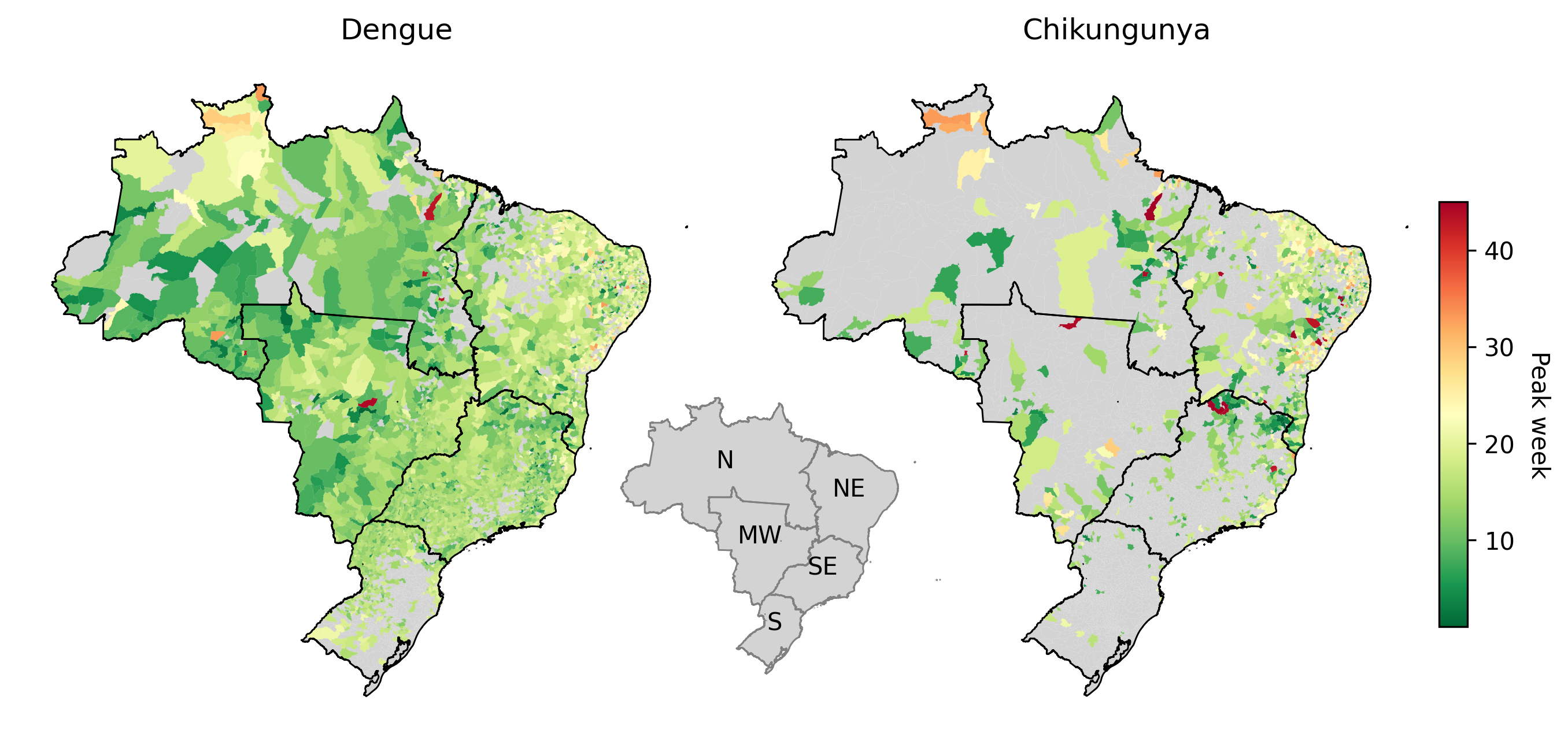}
\caption{Median peak weeks estimated for Dengue and Chikungunya epidemics between 2010 and 2023 by municipality. The inset map shows Brazil's geographic regions: North(N), Northeast (NE), Midwest (MW), Southeast (SE), and South (S).}
\label{fig:map_pw}
\end{figure}

\begin{figure}[!h]
    \includegraphics[width=5in]{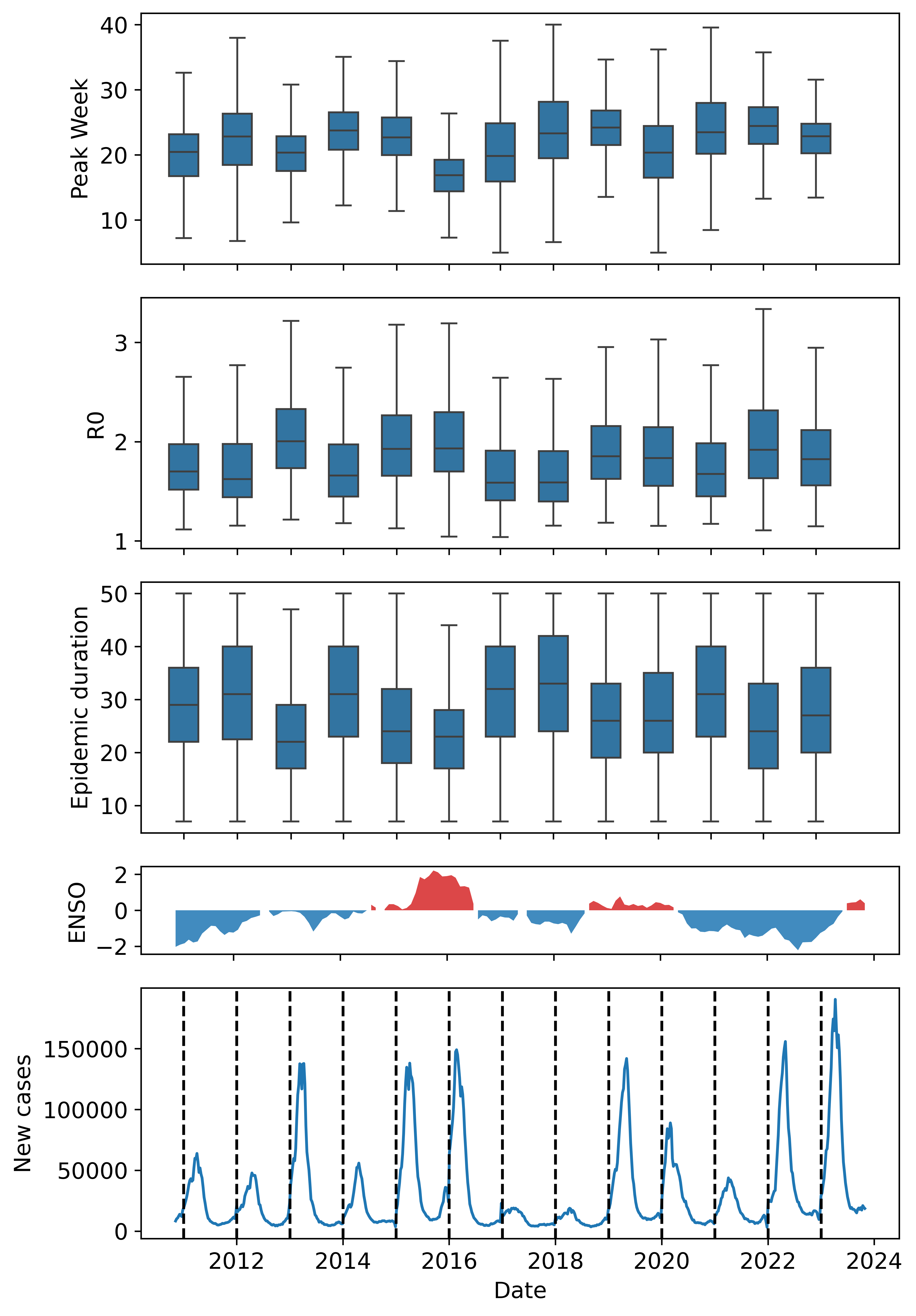}
    \caption{The boxplots, from top to bottom, show the estimated peak week, the estimated basic reproduction number, and the epidemic duration in weeks, for dengue epidemics by year. Above, we find the multivariate ENSO index v2 series, and the last panel is the time series of dengue cases in Brazil between November 2010 and November 2023.}
    \label{fig:time_series}
\end{figure}

The estimated duration of dengue epidemics varied from short ($<10$ weeks) to year-long. Longer epidemics were seen mostly in the North region (Figure \ref{fig:map_duration}). The peak week tended to occur earlier for dengue than chikungunya and in the west earlier than the east side of the country (see Figure \ref{fig:map_pw}). Looking at the country-wide time series for dengue, we can see that higher incidence peaks are associated with shorter epidemics with higher ${\cal R}_0$ (Figure \ref{fig:time_series}). 

A nonlinear relationship was seen between epidemics duration and $R_0$ (see Figure \ref{fig:durationr0br}). Epidemics with high $R_0$ tended to be shorter, while outbreaks with low $R_0$ tended to last longer. This trend was also discernible, albeit less prominently, in the context of chikungunya epidemics, see Figure S2 of the supplementary material.

\begin{figure}[!h]
	\includegraphics[width=5in]{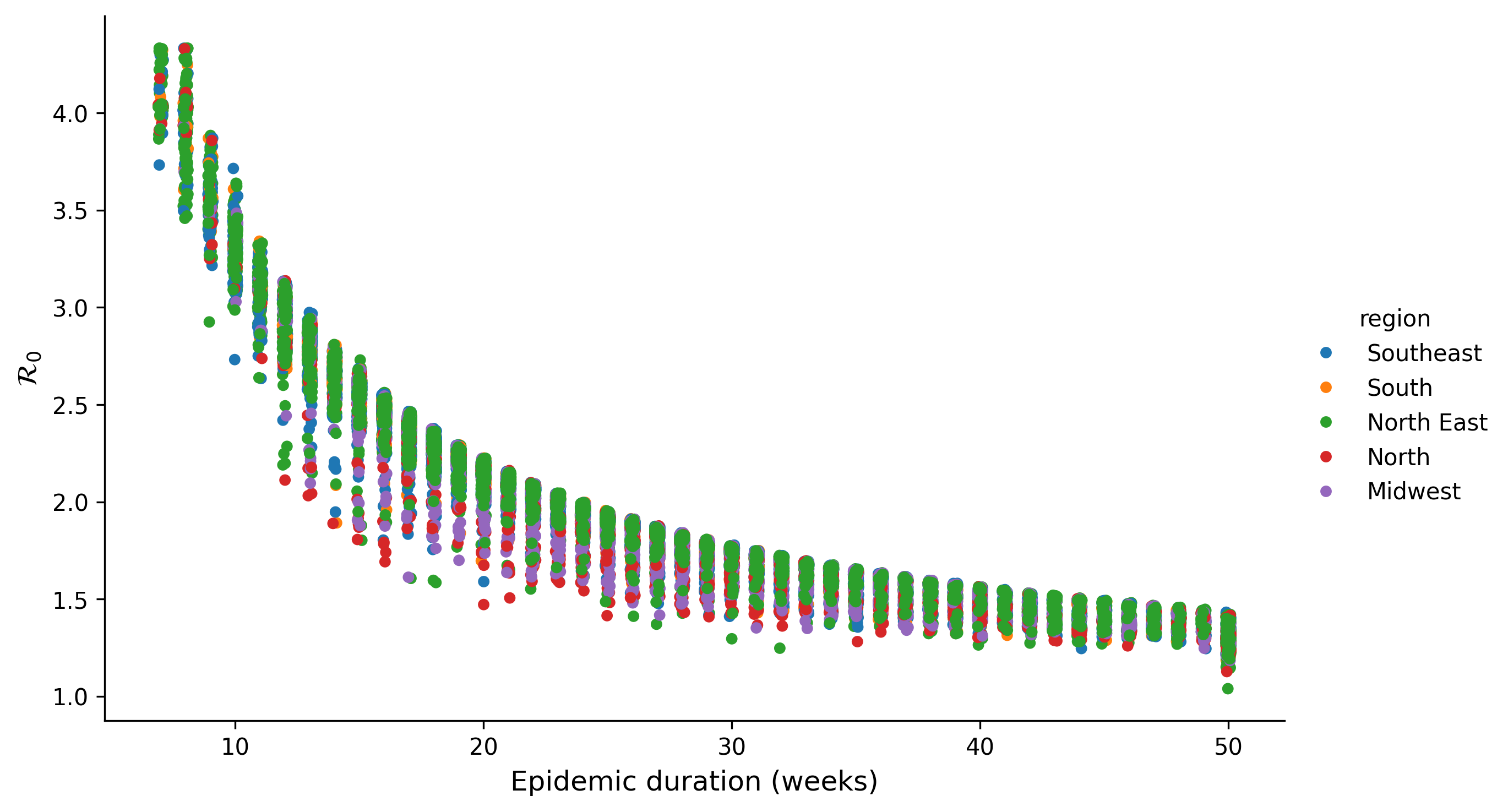}
	\caption{Estimated duration of the dengue epidemic as a function of estimated ${\cal R}_0$.Longer epidemics are associated with lower estimates for R0. Dots are individual cities.}
	\label{fig:durationr0br}
\end{figure}

\subsection{Earlier epidemics and climate}

\begin{figure}[!h]
 \centering
 \includegraphics[width=5in]{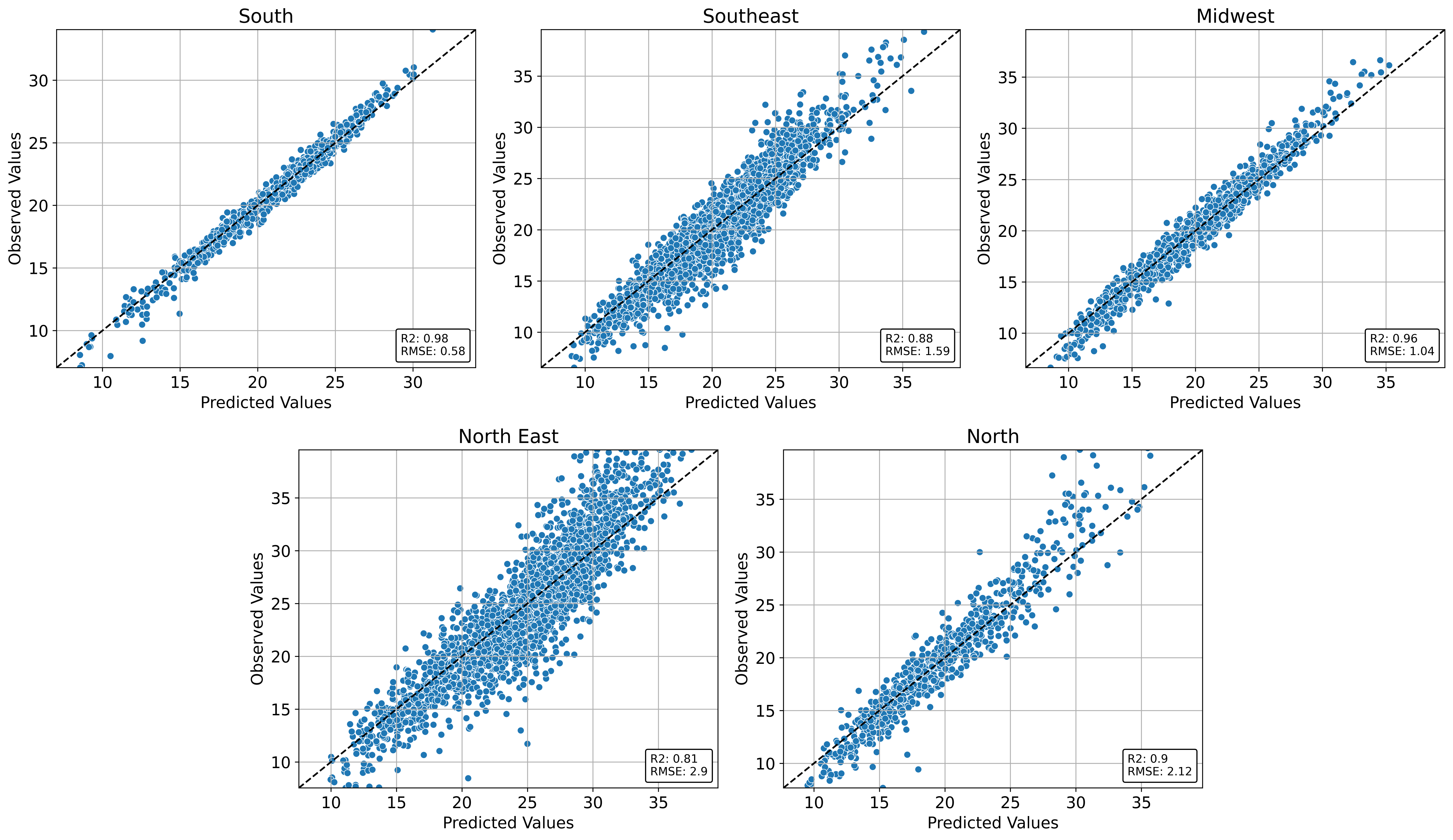}
 \caption{Performance of the Histogram Gradient Boosting regression model trained with the entire dataset. Each point represents the predicted \textit{vs.} observed weeks for a city.  The dashed black line represents the identity.}
 \label{fig:hist_train_model}
\end{figure}

The Figure \ref{fig:hist_train_model} depicts the HGBR model's performance in predicting the dengue epidemic peak week as a function of climate and epidemiological covariates measured one year before. The model provided more accurate predictions in the South, Southeast and Midwest, where seasonality is stronger. 

\begin{figure}[!h]
 \centering
 \includegraphics[width=5in]{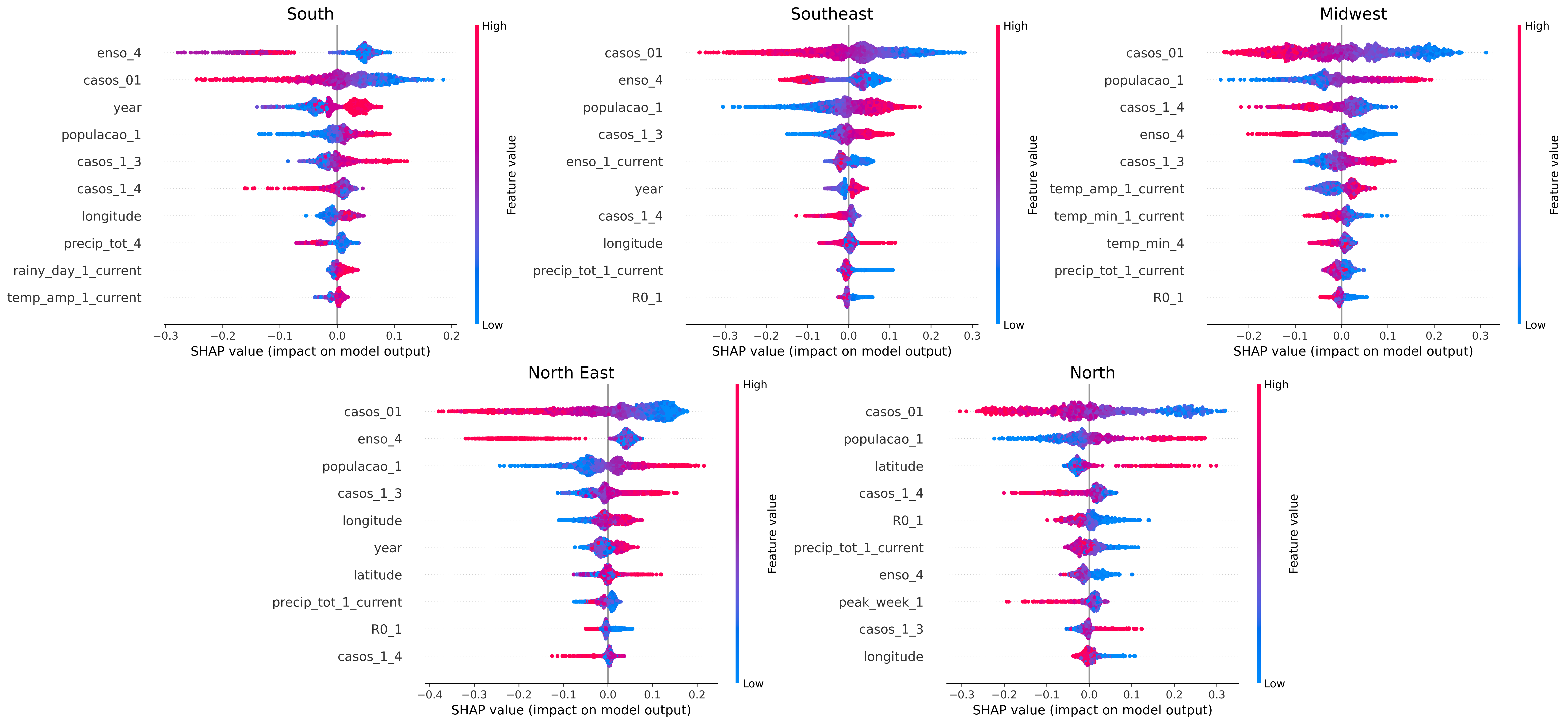}
 \caption{The top 10 most important predictors in each region, encoded as  SHAP values. The SHAP scale represents how much each feature affects the model's output (peak week), with positive values associated with an increase in peak week and negative with a decrease. Each dot represents one epidemic. Their colour is mapped to the value of each feature, with blue standing for lower values and pink for higher. The 0 in the SHAP scale is aligned with the expected value of each feature above.}
 \label{fig:hist_shap_model}
\end{figure}

When we look at the importance of the model's features (Figure \ref{fig:hist_shap_model}), in all regions, the features related to the \textit{peak week}, \textit{population}, and \textit{cases} in the previous year were always among the top ten most important features. Also, all regions had at least two climate predictors in the top ten, with the average temperature in January of the current year being the most common among them.

\begin{figure}[!h]
    \centering
    \includegraphics[width=5in]{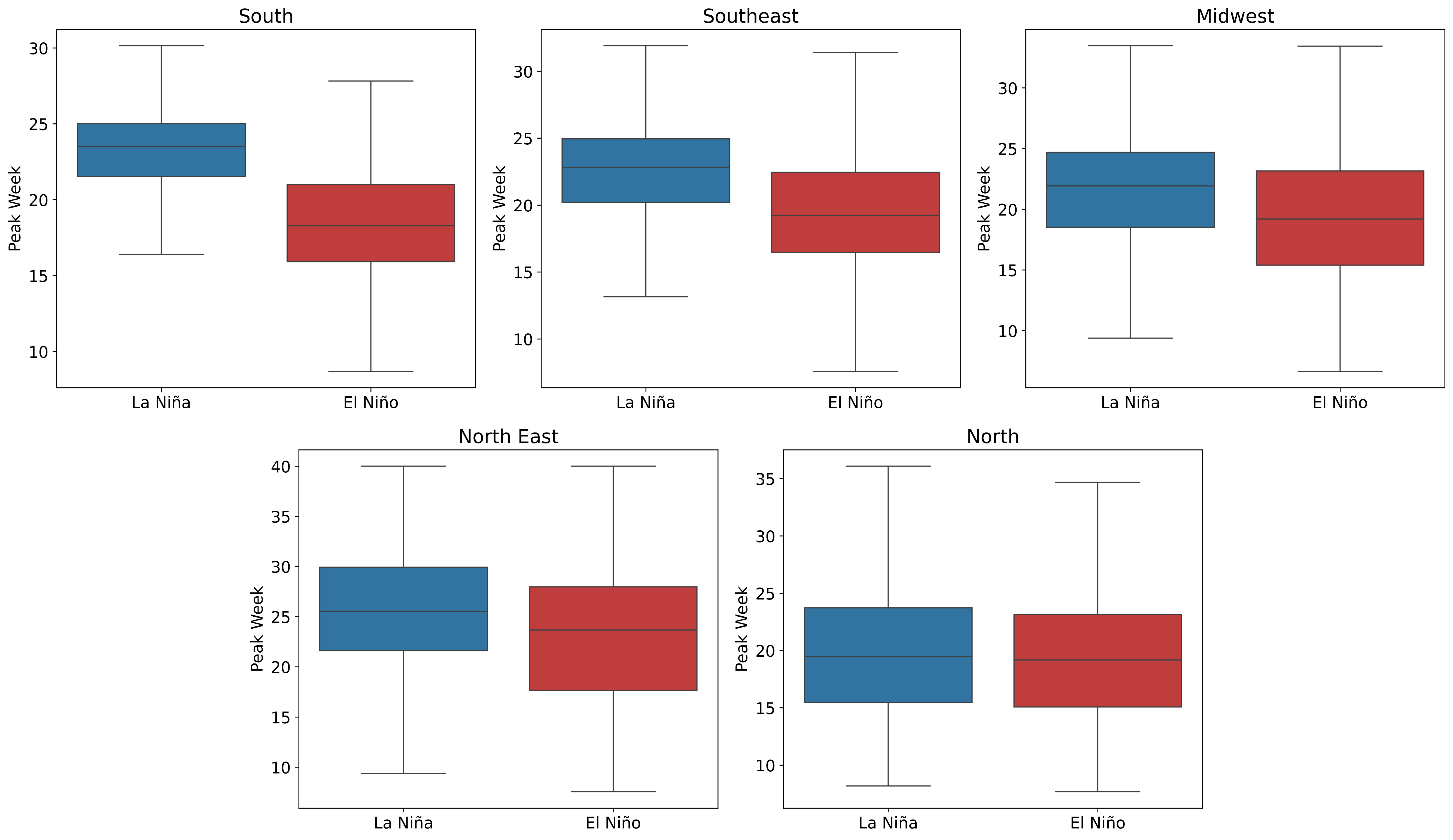}
    \caption{Estimated median peak week of dengue epidemics by region for El Niño and La Niña years. La Niña years, in this analysis, are years when the mean ENSO index in the last quarter of the previous year was below 0, and El niño years are the ones where the mean ENSO index in the last quarter of the year was above 0. }
    \label{fig:box_pw}
\end{figure}

Ocean surface temperature oscillations, as measured by the El niño/La niña index (ENSO), also stood out as an important feature to explain variations in the timing of epidemics. We looked specifically at the value of the ENSO index in the last quarter of the previous year. Positive average ENSO values (El Niño) in that quarter were associated with earlier epidemics in the following year (Figure \ref{fig:box_pw}), while negative 4th quarter ENSO index led to later epidemics. 

\section{Discussion}

This study introduces a computational pipeline~(Figure \ref{fig:workflow}) capable of processing incidence data from all municipalities and estimating key epidemic statistics in a standardized format. To achieve the required computational scalability, we employed optimization techniques to fit the Richards model, which has a one-to-one relationship with the archetypical SIR mathematical model. Even though the SIR model is an approximation to the transmission mechanisms taking place in the field, we believe that this simplified description of the epidemics is a small price to pay in exchange for developing a consistent set of epidemic parameters over a large spatio-temporal range. Using more detailed transmission models would be undesirable, not only because of the added computational cost to fit but also for the added uncertainty associated with the estimation of a larger number of parameters subject to identifiability issues~\citep{gallo2022lack, kao2018practical}.

Following parameter estimation, a regression analysis was conducted to elucidate the influence of local and global climate patterns on epidemic parameters. Although not exhaustive, these exploratory analyses revealed a discernible association between the El Niño Southern Oscillation (ENSO) and epidemic timing (Figure \ref{fig:box_pw}), represented by the peak week of the epidemic. 
The regression analysis results show that various epidemiological, demographic, and climatic factors determine the shape of future epidemics. 
This association has been reported before for Singapore~\citep{earnest2012meteorological}, Thailand~\citep{tipayamongkholgul2009effects}, and other regions of the world~\citep{johansson2009multiyear,moraes2019sazonalidade, banu2015impacts,do2014climatic}. However, our study differentiates itself by the larger spatio-temporal scale of the analysis and for generating an open dataset that enables further explorations and will be continuously updated.

The observed association between ${\cal R}_0$ and epidemic duration, shown in figure \ref{fig:durationr0br}, exhibits remarkable consistency, partly reflecting the interdependence between these parameters within the transmission model. Given the environmental influences driving epidemics, for example, the well-described correlation between transmission and temperature, it is noteworthy that these two parameters remain highly correlated, as per theoretical expectations, that assume constant transmission rates and endogenous determination of duration (by depletion of susceptibles). The small size of the epidemics (relative to population size) suggests that such depletion is not to be expected, except in the hypothesis of a massive underreporting of cases in most cities, which could mask larger attack ratios.

The extensive dataset of epidemiological parameters generated by Episcanner has demonstrated utility in developing predictive models, particularly when integrated with complementary datasets such as climate and demographic series. Although predictive modelling was only superficially explored in this study, its potential for future research is considerable. To our knowledge, this dataset represents a unique large-scale compilation of epidemic parameters estimated using a consistent methodology, enabling robust comparisons. We anticipate that statistical modellers will leverage this open dataset for future investigations.

\section{Conclusion}
The methodological framework introduced in this paper facilitates the comprehensive characterization of epidemic dynamics across both temporal and spatial dimensions. This approach offers invaluable insights into the transmission patterns of arbovirus diseases, enabling public health authorities to devise timely interventions to mitigate disease spread effectively. Furthermore, the identification of spatial variations in epidemic parameters underscores the imperative for region-specific strategies in disease surveillance and control initiatives.

The predictive potential for this parameter set has been demonstrated and can be further elaborated to reveal more specific associations at smaller spatial scales. One limitation of such models is the lack of data on the immunological status of the population, i.e.,  the proportion of the population previously exposed to each dengue serotype. Data about the proportion of cases infected with each DENV type per week or at least per year would contribute significantly to understanding these epidemics' long-term dynamics. We believe that making such data available would be the most important governmental investment in controlling arbovirus diseases.

The epidemiological parameters used in this paper are updated weekly and are made available through Episcanner, a publicly available web dashboard~\citep{Realtime_Epi_Report}.

Continuous monitoring of the epidemiological characteristics (described in section \ref{sec:epi_estimation}) of dengue and chikungunya can be a key tool for efficient control of their incidence and morbidity. When it comes to accurate epidemiological assessments, methodological consistency over time and geographical space is indispensable. In particular, for robust comparisons across time and space, employing the same analytical methods is crucial to ensure that methodological differences are not confounded with data variations.

This study elucidates longer-term association patterns between dengue epidemics' intensity and global climate variations, as exemplified by the multivariate ENSO index. These findings reinforce the longstanding hypothesis~\citep{reiter2001climate} that the ongoing escalation in global average temperatures will inevitably exacerbate the burden of mosquito-borne diseases.

\section*{Acknowledgment}
This work was funded by a grant from the Wellcome Trust (226088/Z/22/Z).

\bibliography{episcanner}

\newpage
\appendix

\setcounter{table}{0}
\setcounter{figure}{0}
\renewcommand{\thetable}{S\arabic{table}}
\renewcommand{\thefigure}{S\arabic{figure}}

\begin{longtable}{m{12cm} |  m{2.5cm}  } 

  \hline
   \textbf{Feature description}  & \textbf{Type} \\ 
  \hline
  \textbf{year:}  year of the peak week  that are being predicted. & temporal \\ 
   \textbf{casos\_01:}   sum of cases in the January of the year whose peak week are being predicted.  & epidemiological \\ 
  \textbf{casos\_1\_3:} sum of cases on the third quarter of the previous year.  &  epidemiological \\ 
  \textbf{casos\_1\_4:} sum of cases on the fourth quarter of the previous year. &  epidemiological \\ 
  \textbf{populacao\_1:} population in the previous year.  &  demographic \\ 
  \textbf{peak\_week\_1:}  peak week estimated on the previous year. &  epidemiological \\ 
  \textbf{R0\_1:} reproduction number estimated on the previous year. &  epidemiological \\ 
  \textbf{ep\_dur\_1:} epidemic duration estimated on the previous year. &  epidemiological \\ 
 \textbf{dummy\_ep:} 1 if the previous year were an epidemic identified by Episcanner and 0 otherwise.  &  epidemiological \\           
 \textbf{temp\_med\_4:} average of the average temperature over the fourth quarter of the previous year. & climatic \\ 
  \textbf{temp\_amp\_4:} average of the temperature amplitude  over the fourth quarter of the previous year. & climatic   \\  
   \textbf{temp\_max\_4:} average of the maximum temperature over the fourth quarter of the previous year.  & climatic \\  
    \textbf{temp\_min\_4:} average of the minimum temperature over the fourth quarter of the previous year. & climatic \\        
 \textbf{umid\_min\_4:} average of the minimum humidity over the fourth quarter of the previous year. & climatic \\    
      \textbf{umid\_max\_4:} average of the maximum humidity over the fourth quarter of the previous year.  & climatic \\      
       \textbf{umid\_amp\_4:}average of the  humidity amplitude over the fourth quarter of the previous year.  & climatic \\      
        \textbf{enso\_4:} average of the  multivariate ENSO (El Niño-Southern Oscillation) in the fourth quarter of the previous year. & climatic \\ 
 \textbf{precip\_tot\_4:} sum of the total precipitation over the fourth quarter of the previous year. & climatic \\ 
  \textbf{rainy\_day\_4:} sum of days with rain (precipitation above zero) over the fourth quarter of the previous year.& climatic \\ 
   \textbf{thr\_temp\_min\_4:} sum of days with minimum temperature below 15-celsius degrees & climatic \\  
    \textbf{thr\_temp\_amp\_4:} sum of days with temperature amplitude above celsius degrees. & climatic \\   
     \textbf{thr\_umid\_med\_4:} sum of days with average humidity above 0.8. & climatic \\         
 \textbf{temp\_med\_1\_current:}  average of the average temperature over the january of the year whose peak week is being predicted.  & climatic \\ 
 \textbf{temp\_amp\_1\_current:} average of the temperature amplitude over the january of the year whose peak week is being predicted.  & climatic \\ 
 \textbf{temp\_max\_1\_current:} average of the maximum temperature over the january of the year whose peak week is being predicted.  & climatic \\ 
 \textbf{temp\_min\_1\_current:} average of the minimum temperature over the january of the year whose peak week is being predicted.  & climatic \\ 
 \textbf{precip\_tot\_1\_current:} sum of total precipitation over the january of the year whose peak week is being predicted.  & climatic \\ 
 \textbf{rainy\_day\_1\_current:}  sum of days with precipitation over the january of the year whose peak week is being predicted.  & climatic \\ 
 \textbf{enso\_1\_current:} average of the ENSO (El Niño-Southern Oscillation) over the January of the year whose peak week is being predicted. & climatic \\ 

 \textbf{latitude:} 
latitude of  the city center. & spatial \\ 
 \textbf{longitude:} 
longitude of the city center. & spatial \\ 
\hline

\caption{Description of the features used in the Histogram Gradient Boosting Regressor model. The climatic data are obtained daily, and the number of cases is obtained weekly. In the features identified with long tails as cases and rainy days, a log transformation was applied to improve the model performance.}

\label{tab:features}

\end{longtable}

\begin{figure}[H]
\centering
	\includegraphics[width=\linewidth]{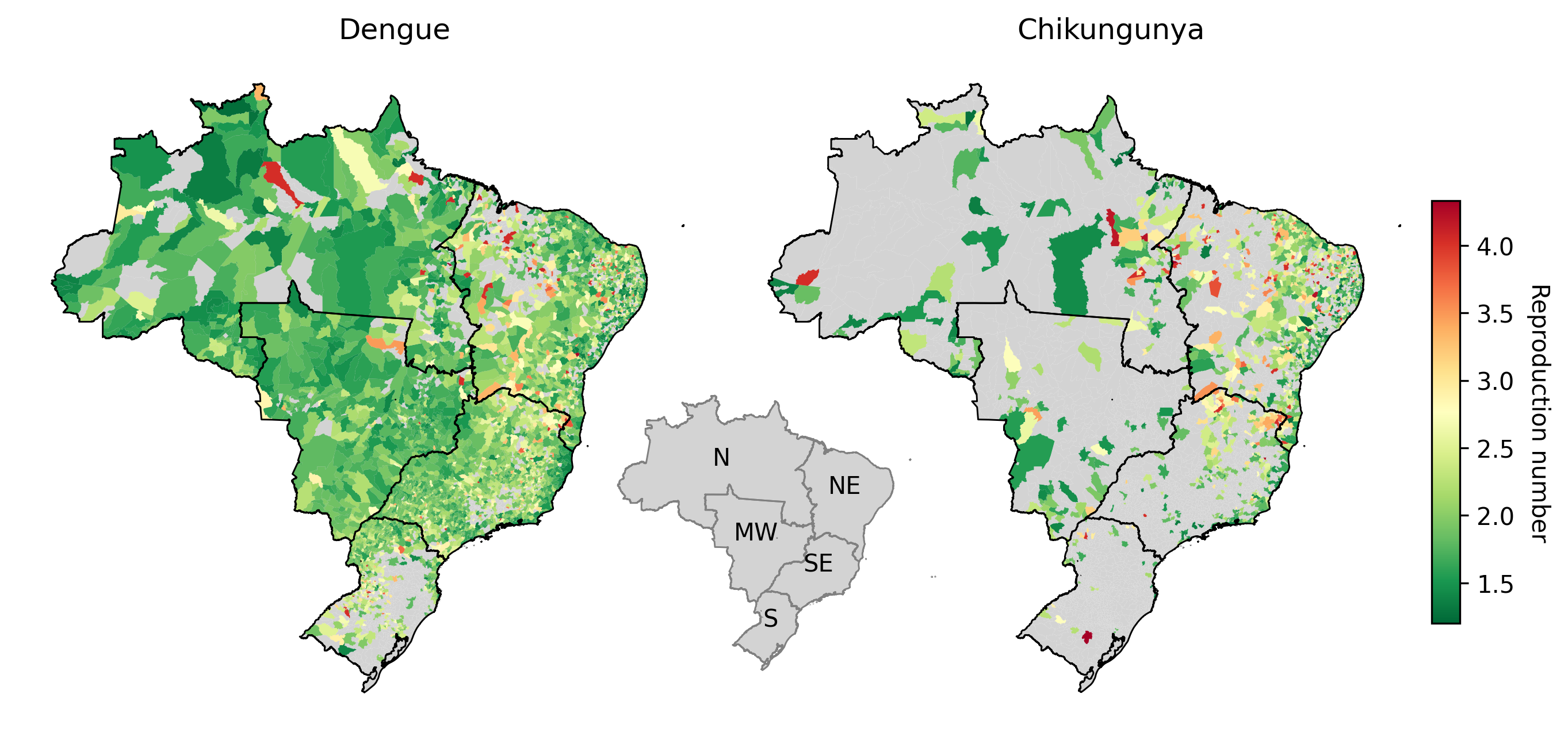}
	\caption{Median of the reproduction number ($R_0$) estimated for Dengue and Chikungunya cases. In the middle map, N refers to the North, NE to the North East, MW to the Midwest, SE to the Southeast, and S to the South.}
	\label{fig:map_r0_dengue_chik}
\end{figure}

\begin{figure}[H]
	\centering
	\includegraphics[width=0.9\linewidth]{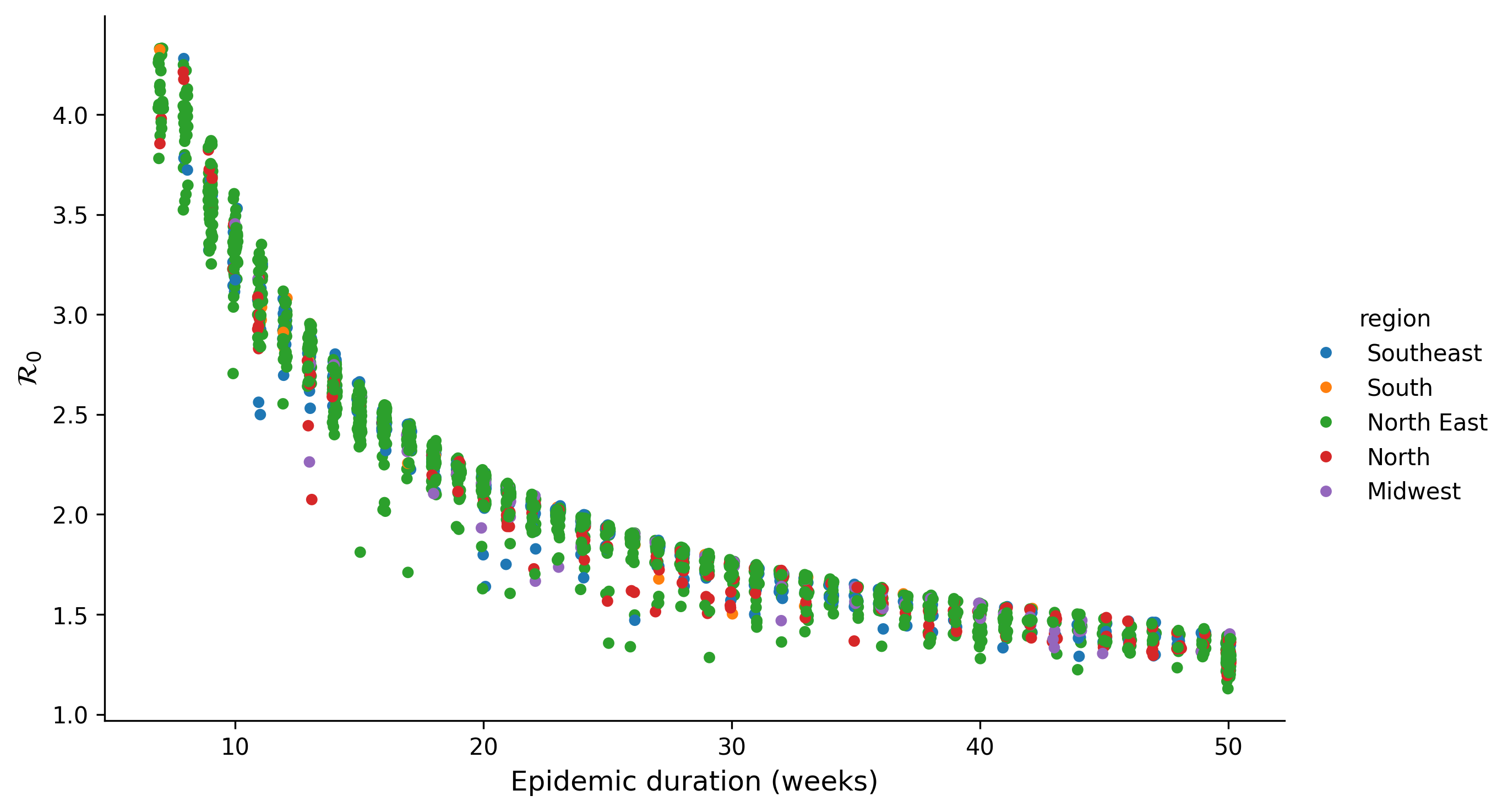}
	\caption{Duration of the chikungunya epidemic as a function of R0. Longer epidemics are associated with lower estimates for R0.}
	\label{fig:durationr0br}
\end{figure}

\end{document}